\documentstyle[12pt]{article}
\newcommand{\lea}{\raisebox{-.3ex}{\small $ \
\stackrel{\textstyle <}{\sim} $ }}
\newcommand{\gea}{\raisebox{-.3ex}{\small $ \
\stackrel{\textstyle >}{\sim} $ }}

\begin{document}

\begin{center}
{\LARGE\bf Comment on ``Phase-shift analysis of NN scattering below
160 MeV: Indication of a strong tensor force''}
\\
\vspace*{.5cm}
{\large\sc G. E. Brown}
\\
{\it Department of Physics, State University of New York,
Stony Brook, New York 11794}
\\
and
\\
{\large\sc R. Machleidt}
\\
{\it Department of Physics, University of Idaho,
 Moscow, ID 83843, U.S.A.}
\\
\today
\end{center}
\vspace*{.5cm}

\begin{abstract}
In his recent publication of a NN phase shift analysis below 160 MeV,
Henneck reports relatively large values for the mixing parameter
$\epsilon_1$. Based on these results, Henneck suggests that the strength
of the $\rho$-meson tensor coupling to the nucleon may be weaker
than used in present day NN interactions, like the Paris or Bonn potentials.
We point out that at low energies ($\lea 100$ MeV)
there is very little sensitivity to the strength of the $\rho$
coupling, due to the compensating effect of the second order
tensor term. In order to establish sensitivity, one has to go
to energies $\gea 200$ MeV, where the second order contribution has
gone out.
As it happens, the $\epsilon_1$ mixing parameter is well determined
in the region of energies 200--300 MeV, and there is agreement with
the predictions by the Paris and Bonn potentials; whereas the weak-$\rho$
model is about 50\% above the data.
This and additional considerations in triplet $P$-waves
 re-confirm that NN scattering requires the
strong $\rho$, consistent with the $\pi\pi-N\bar{N}$
partial-wave analysis by H\"ohler and Pietarinen.
\end{abstract}

Following the measurement of the longitudinal spin correlation
coefficient $A_{zz}$ at 67.5 MeV by the Basel group~\cite{Ham91},
there has been a lot of controversy about the strength of the tensor force
in the nucleon-nucleon (NN) interaction and about the strength
of the $\rho$-meson tensor coupling.

Henneck has published a recent paper: ``Phase-shift analysis of
NN scattering below 160 MeV: Indication of a strong tensor
force''~\cite{Hen93}.
In a fundamental sense, it is unreasonable
to include ``Indication of a strong tensor force'' in a title.
Since half a century,
we all know from the deuteron that there is a strong tensor force.

In both Refs.~\cite{Ham91} and \cite{Hen93}
the implication is that the tensor coupling of the $\rho$-meson
must be weaker than used in present day NN interactions:
Bonn~\cite{MHE87,Mac89} or Paris~\cite{Lac80};
more like the vector-dominance value.

In contradiction to these claims, Klomp {\it et al.}~\cite{Klo92}
and Machleidt and Slaus~\cite{MS93} show from phase shift fits,
that the Bonn and Paris potentials do well in
reproducing the $S$-$D$ mixing parameter $\epsilon_1$,
about which the controversy revolves, cf.~Fig.~1.

Moreover,
a recent paper by Wilburn {\it et al.}~\cite{Wil93}
shows Henneck's value of
$\epsilon_1$, arrived at from his phase shift analysis,
to lie exactly on the Bonn curve at 25 MeV,
and the low-energy measurements of $\epsilon_1$ to lie nicely
(aside from a deviation the authors do not believe to be real)
and to follow well the Bonn curve.

By weak and strong $\rho NN$ coupling we shall mean
$\kappa_\rho = 3.7$ and $\kappa_\rho \approx 6$, respectively;
where $\kappa_\rho\equiv f_\rho/g_\rho$, the ratio of the tensor
to vector coupling constant.
The weak $\rho$-coupling is arrived at by using vector dominance~\cite{Sak69};
the strong $\rho$-coupling is obtained from a $\pi\pi-N\bar{N}$
partial-wave analysis conducted by H\"ohler and Pietarinen~\cite{HP75}.
Later work by Grein~\cite{Gre77}
basically confirmed the H\"ohler-Pietarinen result.

The usual argument about the connection of the strength of the tensor
interaction and that of the $\rho$-meson coupling goes as
follows:

The $\rho$-exchange tensor interaction has opposite sign to that from
pion exchange~\cite{BBN85,Mac89}.
Therefore, summing up the $\rho$-exchange will decrease the tensor force.
However, this argument is too simplistic as was realized long ago.

If the $\rho$-exchange tensor interaction is weak, of the vector dominance
value, then the tensor force will be strong, because not much of the pion
contribution is cancelled (see Fig.~8a of Ref.~\cite{BBN85}).
In this case, second-order effects of the tensor interaction will be
strong [Eq.~(2.10) of Ref.~\cite{BBN85}] and one has an effective
tensor contribution arising from these second order terms
of
\begin{equation}
V_{eff}(r) = \frac{(3-2\mbox{\boldmath $\tau_1 \cdot \tau_2$})}
{\bar{E}} 2 {\bf S}_{12} (V^{(1)}_{tensor})^2
\end{equation}
where the average energy $\bar{E}\approx 200$ MeV.
Note that the {\boldmath $\tau_1 \cdot \tau_2$} piece is negative, of
opposite sign to the $\pi$-exchange tensor. This is, of course, easy to
understand, because in the iterated exchange one has a box diagram,
and the two pions in the crossed channel must, while the
interaction is isovector, be in a $P$-wave because of Bose statistics.
Consequently, a tensor interaction with weak-$\rho$ coupling builds up an
effective $\rho$-meson type coupling, strengthening the $\rho$-channel.
Once second-order effects are included, the net result is little
different for weak- and strong-$\rho$ coupling, although
different amounts of the
effective $\rho$ (iterated pion exchange) result in the two cases.

All potentials, with weak or strong $\rho$-tensor coupling, were constrained
to fit the deuteron. Thus, it is no surprise that Henneck's point
at 25 MeV lies on the Bonn curve~\cite{Wil93}.

Our conclusion is that it is doubtful whether low-energy scattering
experiments distinguish between weak and strong $\rho$-couplings.
In going to higher energies, the second order term, Eq.~(1), will
tend to go out. How high in energy does one have to go?

In order to answer this question,
let us remember that the main part of the symmetry energy in nuclei comes from
the second order tensor interaction
\begin{equation}
V_{symm} \approx + \frac{12}{\bar{E}} (V^{(1)}_{tensor})^2 \; ,
\end{equation}
as one can deduce from Eq.~(2.10) of Ref.~\cite{BBN85}.
Brown, Speth, and Wambach~\cite{BSW81}
showed that the $V_\tau$ so obtained dropped over a scale of
energies 100 to 200 MeV, as the incident nucleon energy
eats into the principal value integral.
Of course, some of the symmetry energy comes from the vector
coupling of the $\rho$-meson, but one can see directly from the work
of Ref.~\cite{BSW81} that this is small; very little
of the interaction drops at the slow rate that $V_{\sigma\tau}$,
which comes mainly from the Born term, drops.

The work of Ref.~\cite{BSW81} employed a strong-$\rho$ tensor
coupling, giving a relatively weak $V^{(1)}_{tensor}$.
The calculations are straightforward, and would have given a
much greater symmetry energy had a weak-$\rho$ coupling been used.

Our first statement, which can be directly tested by calculations,
is that the symmetry energy in nuclear matter calculations will be
much too large if a weak-$\rho$ coupling is used.
Equivalently, the $V_\tau$ used in investigations of isobaric
analog states will be much too large.

Calculations of these matters introduce various many-body
intermediate steps, which lead to suspicion in the views
of experimentalists. However, the work of Ref.~\cite{BSW81} points out
that if one wants to see differences between the weak-$\rho$
and strong-$\rho$ scenarios, one should go to scattering energies
$E > 100$ MeV, by which time some of the second-order contribution of
the tensor interaction is stripped off. Energies of 200 MeV would be better.

Our conclusion is that it would be safest to determine $\epsilon_1$
at energies $\gea 200$ MeV where the second order contributions
from the tensor interaction have been stripped off.

The Henneck $\epsilon_1$ at 50 MeV is about 2.8$^0$, what one would obtain
with zero $\rho$-coupling (cf.\ Fig.~1).
The Reid potential~\cite{Rei68}, with essentially vector
dominance coupling, gives $\approx 2.4^0$, whereas Paris~\cite{Lac80} and
Bonn~\cite{MHE87,Mac89}
(which both use the strong $\rho$)
give $\lea 2^0$. Note that the square of the tensor coupling of the
$\rho$ has increased a factor of more than 2.5 in going from vector
dominance to strong-$\rho$ coupling, with a change of $\approx 0.4^0$ in
$\epsilon_1$. This is what we mean by insensitivity at low energy.
Quoted error
in Henneck is $\pm 0.25^0$.

Since Reid fits the deuteron, we believe that this potential gives
a good indication of what weak-$\rho$ coupling will give in the
200-300 MeV region, once the second-order contribution to the
effective-$\rho$ exchange has gone out. Reid gives
$\epsilon_1 \approx 7^0$ in this region;
while Paris and Bonn predict $\approx 4^0$, in perfect agreement with
the phase shift analyses
by Arndt~\cite{Arn93}, Bugg and Bryan~\cite{BB92}, and by the
Nijmegen group~\cite{Sto93} (cf.\ Fig~1).
This confirms that NN scattering requires the strong $\rho$.

We note that there are phase parameters
which are even more suitable than $\epsilon_1$ to pin down the $\rho$-coupling
strenght,
namely, the the triplet $P$-wave phase shifts.
In contrast to $\epsilon_1$, the $^3P_J$ phase shifts are reliably
determined and there is no controversy among different researchers
conducting phase shift analyses.
Moreover, the effect of the $\rho$-meson is very large in $P$-waves.
For $^3P_0$ and $^3P_2$,
we demonstrate this in Fig.~2, where
the predictions by the weak and the strong $\rho$ are shown.
It is clearly seen that these $P$-waves require by all means
the strong $\rho$.
This is probably the best argument why NN scattering needs the
large $\rho$ tensor coupling, consistent with the determination
by H\"ohler and Pietarinen~\cite{HP75} in 1975.

H\"ohler~\cite{Hoh93} has recently reviewed developments since
the H\"ohler-Pietarinen work.
He points out that a somewhat different method led to
$\kappa_\rho = 6.1 \pm 0.6$~\cite{Pie78}.
Furthermore, the large H\"ohler-Pietarinen
value agrees with an unpublished calculation
by Gustafson, Nielsen, and Oades~\cite{GNO83}.
We mentioned the work by Grein~\cite{Gre77} using NN forward
dispersion relations, which is compatible with H\"ohler-Pietarinen.
The chief point of H\"ohler~\cite{Hoh93} is that the full
information on the $\rho NN$ coupling is contained in the
$\pi\pi-N\bar{N}$ $P$-wave helicity amplitudes and the direct way
to extract coupling constants in an approximate description
was employed in Ref.~\cite{HP75}. These helicity amplitudes
were obtained from $\pi N$ partial wave amplitudes with
imposition of unitarity and analyticity by H\"ohler and Pietarinen.
Work since that time has not consistently enforced
these constraints~\cite{Hoh93}.

In conclusion, we have shown that at low energies
($\lea 100$ MeV)
there is very little sensitivity
to the strength of the $\rho NN$ coupling.
In order to establish sensitivity, one has to go to higher energies
($\gea 200$ MeV).
Phase shift analyses here clearly
favor the strong $\rho NN$ coupling.
Independently of this empirical argument,
the strong $\rho NN$ coupling was firmly established in 1975
from detailed knowledge of the
$\pi\pi-N\bar{N}$ helicity amplitudes~\cite{HP75}.

\vspace*{1cm}
This work was supported in part by the U.S. National Science Foundation
under Grant No.~PHY-9211607 and by the U.S. Department of Energy
under Grant No.~DE-FG02-88ER40388.

\pagebreak

\begin{center}
\large\bf
Figure Captions
\end{center}

\noindent
{\bf Figure 1.} The $\epsilon_1$ mixing parameter at low and
intermediate energies.
Predictions by the Bonn-B~\cite{Mac89} (solid line),
Paris~\cite{Lac80} (solid line), and
Reid potential~\cite{Rei68} (dashed),
 as well as a model that does not include
a $\rho$ meson (`No $\rho$', dotted line) are shown.
The phase shift analysis by Henneck~\cite{Hen93} is represented by
the diamonds. Furthermore,  the analyses by
Arndt~\cite{Arn93} (solid triangles),
Bugg and Bryan~\cite{BB92} (solid dots),
and the Nijmegen group~\cite{Sto93} (solid squares)
are displayed.

\vspace*{.5cm}
\noindent
{\bf Figure 2.} The $^3P_0$ and $^3P_2$ phase shifts of
proton-proton scattering.
The solid line gives the prediction by a meson model that includes
the strong $\rho$, while the dashed line is obtained using the
weak $\rho$.
The solid dots represent the Nijmegen $pp$ multi-energy phase
shift analysis~\cite{Sto93}.
(From Ref.~\cite{Mac93}.)


\vspace*{2cm}

{\it The figures
are available upon request from
\begin{center}
{\sc machleid@tamaluit.phys.uidaho.edu}
\end{center}
Please include your FAX number or mailing address with your request.}


\begin{thebibliography}{99}
\bibitem{Ham91} M. Hammans {\it et al.}, Phys. Rev. Lett. {\bf 66},
2293 (1991).
\bibitem{Hen93} R. Henneck, Phys. Rev. C {\bf 47}, 1859 (1993).
\bibitem{MHE87} R. Machleidt, K. Holinde, and C. Elster, Phys. Reports
{\bf 149}, 1 (1987).
\bibitem{Mac89} R. Machleidt, Adv. Nucl. Phys. {\bf 19}, 189 (1989).
\bibitem{Lac80} M. Lacombe {\it et al.}, Phys. Rec. C {\bf 21},
861 (1980).
\bibitem{Klo92} R. A. M. Klomp, V. G. J. Stoks, and J. J. de Swart,
Phys. Rev. C {\bf 45},
2023 (1992).
\bibitem{MS93} R. Machleidt and I. Slaus, {\it Comment on ``Neutron-Proton
Spin-Correlation Parameter $A_{zz}$ at 68 MeV''}, submitted to
Phys. Rev. Lett.
\bibitem{Rei68} R. V. Reid, Ann. of Phys. (N.Y.) {\bf 50}, 411 (1968).
\bibitem{Arn93} R. A. Arndt, NN phase shift analysis of Spring 1993
(SP93), SAID.
\bibitem{BB92} D. V. Bugg and R. A. Bryan, Nucl. Phys. {\bf A540},
449 (1992).
\bibitem{Sto93} V. Stoks, R. Timmermans, and J. J. de Swart,
Phys. Rev. C {\bf 47}, 512 (1993).
\bibitem{Wil93} W. S. Wilburn {\it et al.},
Phys. Rev. Lett. {\bf 71}, 1982 (1993).
\bibitem{Sak69} J. J. Sakurai, {\it Currents and Mesons} (University
of Chicago Press, Chicago, 1969).
\bibitem{HP75} G. H\"ohler and E. Pietarinen, Nucl. Phys. {\bf B95},
210 (1975).
\bibitem{Gre77} W. Grein, Nucl. Phys. {\bf B131}, 255 (1977).
\bibitem{BBN85} S.-O. B\"ackman, G. E. Brown, and J. A. Niskanen,
Phys. Reports {\bf 124}, 1 (1985).
\bibitem{BSW81} G. E. Brown, J. Speth, and J. Wambach, Phys. Rev.
Lett. {\bf 46}, 1057 (1981).
\bibitem{Mac93} R. Machleidt, {\it Constraints on the $\pi NN$ coupling
from the NN system}, Invited Talk presented at the 5-th
International Symposium on Meson-Nucleon Physics and the Structure
of the Nucleon, Boulder, Colorado, September 1993, $\pi N$ Newsletter No.~9,
to be published.
\bibitem{Hoh93} G. H\"ohler, {\it What can be learned from elastic
electron-nucleon scattering experiments?},
Invited Talk presented at the 5-th
International Symposium on Meson-Nucleon Physics and the Structure
of the Nucleon, Boulder, Colorado, September 1993, $\pi N$ Newsletter No.~9,
to be published.
\bibitem{Pie78} E. Pietarinen, Helsinki preprint HU-TFT-78-13 (1978).
\bibitem{GNO83} See pp. 228, 271 in G. H\"ohler, {\it Pion-Nucleon
Scattering}, Landolt-B\"ornstein Vol. I/9b, ed. H. Schopper
(Springer, New York, 1983).
\end{thebibliography}
\end{document}